\documentclass[11pt]{article}
\usepackage[margin=2cm]{geometry}
\usepackage[utf8]{inputenc}
\usepackage[backref=page]{hyperref}
\usepackage[round]{natbib}
\usepackage[dvipsnames]{xcolor}
\usepackage{amsmath,amsthm,amssymb,nicefrac}
\usepackage{tikz, algorithm, algpseudocode}
\usepackage{csquotes, booktabs, url, float, caption, tabularx, multirow}

\def\11{{\mathbf 1}}    

\usepackage{xr}
\usepackage{glossaries}
\usepackage{authblk}

\glsdisablehyper

\begin{document}

\newacronym{aerchemmip}{AerChemMIP}{Aerosol Chemistry MIP}
\newacronym{cs3}{CS3}{MIT Center for Sustainability Science and Strategy}
\newacronym{cmip}{CMIP}{Coupled Model Intercomparison Project}
\newacronym{cmip7}{CMIP7}{the seventh phase of the Coupled Model Intercomparison Project}
\newacronym{damip}{DAMIP}{Detection and Attribution MIP}
\newacronym{fair}{FaIR}{Finite amplitude Impulse Response}
\newacronym[plural={EMICs}, longplural={Earth system models of intermediate complexity}]{emic}{EMIC}{Earth system model of intermediate complexity}
\newacronym{esm}{ESM}{Earth system model}
\newacronym{geomip}{GeoMIP}{Geoengineering MIP}
\newacronym{gmst}{GMST}{global mean surface temperature}
\newacronym{ml}{ML}{machine learning}
\newacronym{mesm}{MESM}{MIT Earth System Model}
\newacronym{nrmse}{NRMSE}{normalized root mean squared error}
\newacronym{ramip}{RAMIP}{Regional Aerosol MIP}
\newacronym[plural={RCPs}, longplural={Representative Concentration Pathways}]{rcp}{RCP}{Representative Concentration Pathway}
\newacronym{scm}{SCM}{simple climate model}
\newacronym{ssp}{SSP}{Shared Socioeconomic Pathway}
\newacronym{zecmip}{ZECMIP}{Zero Emissions Commitment MIP}

\title{Optimal scenario design for climate emulation}

\author[a,b]{Christopher B. Womack}
\author[c]{Shahine Bouabid}
\author[b]{Andrei Sokolov}
\author[b]{Popat Salunke}
\author[c]{Glenn Flierl}
\author[d]{Sebastian D. Eastham}
\author[b,c,e]{Noelle E. Selin}

\affil[a]{Department of Aeronautics and Astronautics, \authorcr Massachusetts Institute of Technology, Cambridge, MA, United States}
\affil[b]{Center for Sustainability Science and Strategy, \authorcr Massachusetts Institute of Technology, Cambridge, MA, United States}
\affil[c]{Department of Earth, Atmospheric, and Planetary Sciences, \authorcr Massachusetts Institute of Technology, Cambridge, MA, United States}
\affil[d]{Brahmal Vasudevan Institute for Sustainable Aviation, Department of Aeronautics, \authorcr Imperial College London, London, United Kingdom}
\affil[e]{Institute for Data, Systems, and Society, \authorcr Massachusetts Institute of Technology, Cambridge, MA, United States}

\date{}

\maketitle

\begin{abstract}
    As deep learning for physical systems continues to grow in popularity, efforts to improve generalizability have primarily focused on designing architectures that embed physical constraints. However, for machine-learning surrogate climate models (emulators), we show that the low structural diversity in existing scenarios commonly used to generate training data places a ceiling on predictive skill. Here, we examine whether training datasets themselves can be optimized to improve generalization. We introduce a method to create datasets that produce emulators capable of generalizing to new, structurally different scenarios absent from the training data. We use a differentiable simple climate model (SCM) to calculate the sensitivity of emulator loss to perturbations in the training data, iteratively updating the training data to maximize emulator skill. For an SCM, an emulator trained on one scenario optimized in this fashion outperforms an emulator trained on six standard ScenarioMIP pathways. We achieve this higher predictive skill despite training on a smaller dataset, finding that our emulator successfully isolates distinct physical behaviors of different climate forcing agents (e.g., greenhouse gases vs. aerosols) without single-forcing runs. We then demonstrate that scenarios optimized using an SCM, when used to drive an intermediate-complexity climate model, produce a training dataset that yields a more skillful emulator than training on ScenarioMIP outputs. Our results suggest that, in the compute-constrained environment of running full-scale climate models, generating a small number of dynamically rich scenarios provides greater marginal value for emulation and characterizing system responses than expanding the suite of traditional emissions pathways.
\end{abstract}

\section*{Introduction}

While machine learning (ML) models exhibit immense utility in interpolating complex physical systems, their ability to generalize to unseen, out-of-distribution scenarios while adhering to physical laws remains a fundamental challenge. Efforts to enforce physical consistency typically focus on model architecture and include Physics-Informed Neural Networks (PINNs) that embed governing equations into the loss function \cite{raissi_physicsinformed_2019, cai_physicsinformed_2021, karniadakis_physicsinformed_2021, cuomo_scientific_2022}, operator learning approaches that map directly between function spaces \cite{li_fourier_2021, lu_learning_2021}, and hard constraints (e.g., enforcing conservation or symmetries) \cite{greydanus_hamiltonian_2019, mohan_embedding_2020, satorras_equivariant_2021}. Hybrid techniques such as NeuralGCM further demonstrate that combining physical and statistical components can achieve significant computational savings without sacrificing predictive skill \cite{kochkov_neural_2024, bracco_machine_2024}.

Beyond architectural constraints, the design of the training data themselves dictates whether an ML model learns the underlying physics or interpolates between observed states. Data design methods include physics-informed feature engineering (e.g., using nondimensional quantities such as the Reynolds number instead of raw velocity fields) \cite{fazliani_enhancing_2025}, physics-guided data augmentation that exploits known invariances or linearity properties \cite{li_physicsguided_2022}, and synthetic data generation via active learning to place new samples in regions of large physical error or high model uncertainty \cite{shields_active_2023, guo_active_2024}. Such methods may be particularly impactful in climate science, as the high computational cost of large-scale simulations restricts the availability of training data \cite{balaji_cpmip_2017, keller_replicability_2025}. 

In climate science, ML emulators address the demand for spatially explicit projections beyond the standard suite of plausible emissions scenarios simulated as part of the \gls{cmip} \cite{eyring_overview_2016,  vanvuuren_scenario_2026}. Following Tebaldi et al. \cite{tebaldi_emulators_2025}, we define emulators as statistical surrogates for physical models, distinct from process-based \glspl{scm} and \glspl{emic}. Reliable climate projections are crucial for areas such as agriculture \cite{hultgren_impacts_2025}, the built environment \cite{crawley_estimating_2008}, energy systems \cite{yalew_impacts_2020}, and the finance and insurance sectors \cite{collier_climate_2021, zhou_review_2023}, all of which face substantial physical and transition risks from climate change. Emulators have demonstrated skill in reproducing variables such as near-surface air temperature, precipitation, relative humidity, and wind speed across annual, monthly, and daily timescales \cite{meinshausen_emulating_2011, castruccio_statistical_2014, beusch_emulating_2020, bassetti_diffesm_2024, bouabid_fairgp_2024, bouabid_scorebased_2026, tebaldi_emulators_2025, womack_rapid_2025, bouabid_evidence_2026}.

\begin{figure}[t]
    \centering
    \includegraphics[width=\linewidth]{Figures/fig01_optimization.pdf}
    \caption{Overview of the training data optimization process for an emulator that maps from emissions (input) to global mean surface temperature (output). We iteratively update the input emissions pathways to maximize emulator skill through four steps: (1) train a base emulator on an initial emissions trajectory; (2) test predictive skill on target scenarios; (3) compute the sensitivity of the test loss to the training data via automatic differentiation; and (4) update the training data via stochastic gradient descent. We repeat this until convergence, performing a final independent evaluation on held-out datasets. For more details on this procedure and emulator architecture, see SI Appendices 1 and 3, respectively.}
    \label{fig:01_optimization}
\end{figure}

Assessing whether emulators respect physical constraints remains challenging, as demonstrating physical consistency requires extrapolating to emissions trajectories distinct from those seen in training. In practice, however, most studies emphasize in-sample and within-range performance---where \gls{gmst} or emissions trajectories lie within the training range---with limited emphasis on structurally out-of-distribution tests \cite{watson-parris_climatebench_2022, lutjens_impact_2025, tebaldi_emulators_2025, schongart_review_2026}. This gap stems from the high temporal and computational costs of running full-scale \glspl{esm}, typically limiting emulator developers to the data made available via \gls{cmip} for training and evaluation. As a result, emulators are largely trained on aggregate emissions pathways, such as ScenarioMIP \cite{beusch_emulating_2020, tebaldi_stitches_2022, bouabid_fairgp_2024, geogdzhayev_eofbased_2026, mathison_rapidapplication_2025}. Previous work demonstrates that training on ScenarioMIP-like pathways is not necessarily optimal \cite{womack_theoretical_2026}, as it restricts our ability to test and train emulators that accurately respond to emissions of individual forcing agents (e.g., anthropogenic greenhouse gases and aerosols). This shortcoming is particularly pressing given that \gls{esm} scenario design for future \gls{cmip} efforts is moving toward a broader set of forcing combinations \cite{vanvuuren_scenario_2026}. One solution is to run the \gls{esm} for each individual forcing to generate a broader set of training data \cite{tebaldi_emulators_2025, vankatwyk_rewiring_2026}, but high simulation costs and the potential for nonlinear interactions when combining forcings currently impede both the exploration and adoption of this approach. Consequently, there is a need for an approach that yields highly informative training data at a low computational cost.

Here, we introduce a method to generate optimal emissions scenarios that improve both overall emulator performance and the ability to emulate the climate response to individual forcing agents. By framing training data generation as a problem of optimal experimental design \cite{fedorov_optimal_2010}, we directly optimize the emissions scenarios themselves to maximize emulator predictive skill; high-level and detailed descriptions of this procedure are given in the following section and SI Appendix 1, respectively. Leveraging a differentiable model based on the \gls{fair} \gls{scm} \cite{leach_fairv200_2021}, our approach calculates the sensitivity of emulator predictive skill with respect to the training data, enabling iterative updates of the training data to minimize a user-defined error metric (Fig. \ref{fig:01_optimization}). Using simple and intermediate-complexity climate models as proxies for \gls{esm}-simulated data, we demonstrate that an emulator trained on a single optimized scenario outperforms a baseline emulator trained on a suite of six standard socioeconomic scenarios (ScenarioMIP-CMIP7). Furthermore, the optimized training data yield increases in emulator skill when extrapolating to structurally out-of-distribution scenarios, indicating a more robust statistical mapping from emissions to temperature. We validate the scalability of our approach by running \gls{scm}-optimized scenarios with the \gls{mesm} \cite{sokolov_description_2018}, a zonally averaged \gls{emic}, demonstrating that this method is transferable to models of higher complexity. Finally, we discuss implications for designing \gls{esm} scenarios specifically for emulator training, along with the potential for extending our approach to other data-constrained domains of machine learning for physical systems.

\begin{figure}[t]
    \centering
    \includegraphics[width=\linewidth]{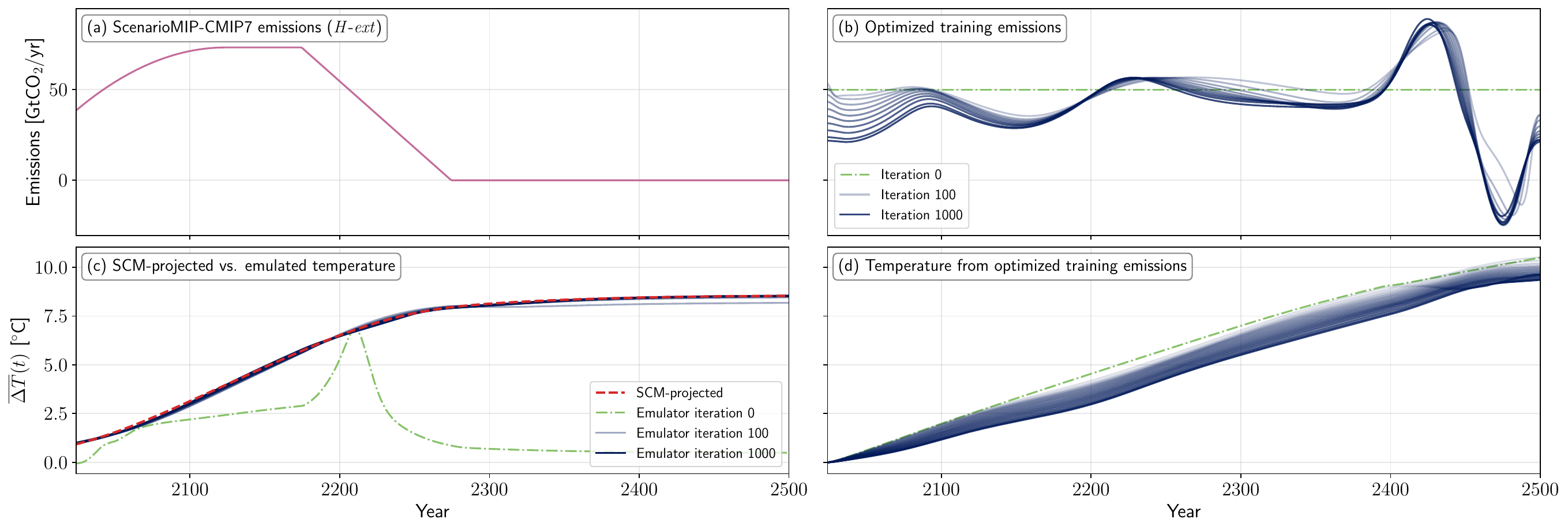}
    \caption{Illustrative optimization results for a single CO$_2$-only high-warming scenario (ScenarioMIP-CMIP7: \textit{H-ext}). (a) Ground-truth emissions trajectory. (b) Evolution of optimized training emissions over 1000 iterations, beginning from a constant initial guess (green dot-dash line). (c) Comparison of SCM-projected (dashed red line) vs. emulated \gls{gmst} predictions (green dot-dash and solid blue lines). (d) Temperature trajectories corresponding to the emissions in (b). Faint lines trace intermediate states every 100 iterations in (b)--(d). The emulator is trained on the synthetic input-output pair (b, d) and tested by predicting the response to ground-truth input (a), as shown in (c).}
    \label{fig:02_co2_example}
\end{figure}

{\setlength{\tabcolsep}{5pt}
\begin{table}[b]
    \centering
    \footnotesize
    \caption{Summary of the experimental protocol utilized in this work. For each climate model, we train a baseline emulator, along with multiple optimized emulator configurations as described in the optimization column.}
    \label{tab:experiments}
    \begin{tabular}{p{0.12\linewidth} p{0.15\linewidth} p{0.3\linewidth} p{0.24\linewidth} p{0.11\linewidth}}
    \textbf{Climate model} & \textbf{Baseline scenarios} & \textbf{Optimization (training data generation)} & \textbf{Evaluation scenarios} & \textbf{Emulator targets} \\
    \midrule
    \textbf{Differentiable SCM} & ScenarioMIP-CMIP7 Priority 1 & Iteratively updated to maximize predictive skill when tested on:\newline 1. Individual sets (ScenarioMIP-CMIP7, DECK, CS3, DAMIP, and GeoMIP)\newline 2. All scenario sets simultaneously & Evaluated against all individual scenario sets & Global Mean Surface Temperature (GMST) \\
    \addlinespace
    \textbf{EMIC (MESM)} & ScenarioMIP-CMIP7 Priority 1 & \textit{No direct optimization.} Uses the optimal emissions trajectories generated by the SCM optimized for performance over all scenarios & Independent evaluation across all single-forcing scenario sets (ScenarioMIP-CMIP7, DECK, CS3) & zonally averaged Temperatures \\
    \end{tabular}
\end{table}}

\section*{Results}

We compare the performance of two emulator configurations: a baseline emulator trained on ScenarioMIP-CMIP7 Priority 1, and an emulator trained on optimized data (hereafter referred to as the optimized emulator). As this work focuses on the impact of training data on predictive skill rather than emulator architecture (i.e., emulator structure and feature design), both configurations use a multi-layer perceptron, the simplest possible neural network emulator. The emulator predicts temperature time series resulting from input emissions trajectories. As the optimization and emulator training procedures are stochastic, we report all results as the median and interquartile range (IQR, 25--75\%) over 50 independent runs. 

We generate optimized training data through a four-part iterative procedure (Fig. \ref{fig:01_optimization}) that treats the training trajectory as a set of tunable parameters (see SI Appendix 1 for a technical description). First, we simulate the temperature response to an initial emissions time series to train a base version of our emulator (Fig. \ref{fig:01_optimization}.1). We refer to this starting trajectory as the initial guess, distinct from the initial conditions of a climate model's physical state. Second, we test the emulator's performance by measuring its predictive skill in terms of \gls{nrmse} over a fixed test dataset (e.g., ScenarioMIP-CMIP7 Priority 1, Fig. \ref{fig:01_optimization}.2); skill is normalized by maximum scenario \gls{gmst} to avoid overemphasizing performance on high-warming scenarios (see Equation 7 in SI Appendix 1.2). Third, we use automatic differentiation to backpropagate through the testing, training, and data generation processes to calculate the sensitivity of the test error to perturbations in the training data (Fig. \ref{fig:01_optimization}.3). Finally, we use stochastic gradient descent to iteratively update the training emissions trajectory to maximize performance (Fig. \ref{fig:01_optimization}.4). In this context, `optimizing for a scenario' strictly means iteratively updating a training emissions trajectory to maximize the resulting emulator's ability to accurately reproduce the temperature response of that specific target scenario (or set of scenarios). Each optimization run updates a single emissions trajectory of fixed length; where we report training on two optimized scenarios, these are produced by separate runs rather than optimized jointly. The optimization procedure is nearly unconstrained, allowing for negative emissions, though we include trajectory smoothness (i.e., relative proportion of high-frequency and low-frequency variations) as a tunable hyperparameter; a full list of hyperparameters and their selection process can be found in SI Appendix 3. To calculate the sensitivity of the emulator error with respect to the training data, we implement a differentiable \gls{scm} (SI Appendix 2) based on the \gls{fair} \gls{scm} \cite{leach_fairv200_2021}, which includes a subset of anthropogenic forcing agents (CO$_2$, CH$_4$, N$_2$O, sulfur, and black carbon); this limited set allows us to focus on the dominant drivers of future warming while retaining a tractable parameter space.

We evaluate the emulators' ability to reproduce temperature anomalies predicted by an \gls{scm} and an \gls{emic} under individual (e.g., CO$_2$-only) and combined forcing scenarios. To test the emulators' performance across different dynamical regimes, we evaluate them against several sets of emissions scenarios. These include plausible future socioeconomic policy projections (the proposed ScenarioMIP-CMIP7\footnote{At the time of performing this investigation and writing this manuscript, the final version of ScenarioMIP-CMIP7 was not yet published. As a result, we use the scenarios outlined in the preprint manuscript, not including the \textit{High-to-Low} scenario added in the final version.} Priority 1 and 2 protocol \cite{vanvuuren_scenario_2026}, and the 2025 MIT Global Change Outlook \cite{paltsev_2025_2025}), along with idealized experiments from the \gls{cmip} DECK designed to display model feedback response characteristics \cite{eyring_overview_2016}.

\begin{figure}[t]
    \centering
    \includegraphics[width=\linewidth]{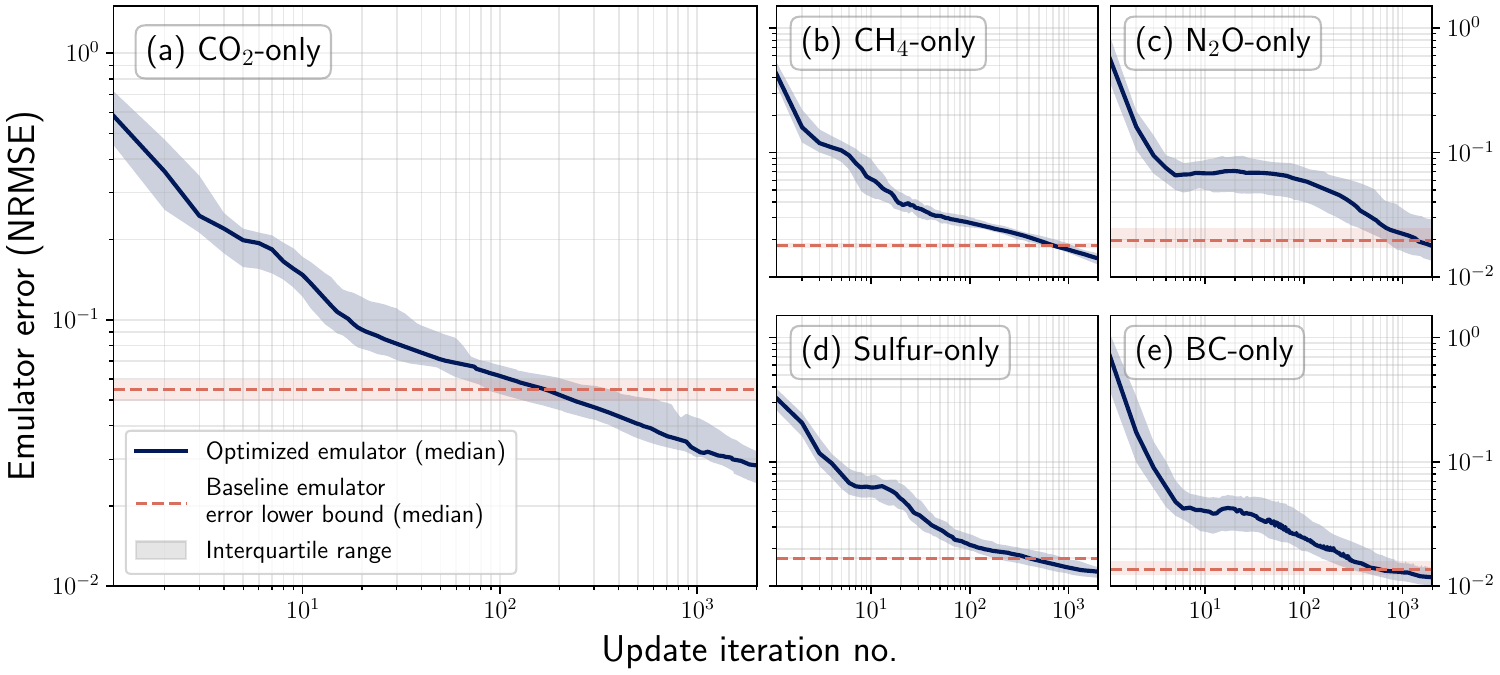}
    \caption{Error spread in emulating single-forcing experiments. Evolution of evaluation loss (\gls{nrmse}) when reproducing \gls{scm}-projected \gls{gmst} anomalies for ScenarioMIP-CMIP7 Priority 1 single-forcing scenarios (e.g., (a) CO$_2$-only, (b) CH$_4$-only). Emulators are evaluated across 50 independent stochastic runs. The solid dark blue line tracks the optimized emulator's median performance, while the dashed orange line indicates the baseline emulator's median error lower bound (evaluated on its own training data). Shaded regions represent the interquartile range (25th to 75th percentiles) of the emulator errors. Full numerical results are provided in SI Appendix 7.}
    \label{fig:03_single_forcing}
\end{figure}

A key feature of a physically consistent emulator is that it can be applied to scenarios whose forcing structure differs from the pathways seen during training and, in our case, optimization. We therefore evaluate an emulator trained to reproduce the effects of multiple forcing agents against two classes of scenarios: those that isolate or decompose the response to individual forcing agents (\gls{aerchemmip} \cite{collins_aerchemmip_2017}, \gls{damip} \cite{gillett_detection_2016, gillett_detection_2025}, \gls{geomip} \cite{kravitz_geoengineering_2015, visioni_geoengineering_2026}, \gls{ramip} \cite{wilcox_regional_2023}, and \gls{zecmip} \cite{jones_zero_2019}), and alternate plausible multi-agent pathways (\glspl{rcp} 4.5 and 8.5 \cite{vanvuuren_representative_2011}, and the \glspl{ssp} \cite{oneill_scenario_2016}). Because our \gls{scm} represents sulfur's radiative contribution through parameterized aerosol-cloud and aerosol-radiation interactions that assume short-lived tropospheric emissions \cite{leach_fairv200_2021}, reproducing the cooling specified by the \gls{geomip} stratospheric injection protocol requires sulfur emissions far exceeding the original specifications; the resulting analog is therefore not a realistic intervention pathway but a stress test at the extreme end of the aerosol forcing range. Similarly, because our \gls{scm} is global, the \gls{ramip} scenarios are constructed as global-average analogs that decouple GHG and aerosol pathways. We do not test regional heterogeneity, but rather a GHG-aerosol ratio that standard pathways do not produce. To enable independent evaluation without risk of information leakage (i.e., evaluation data present in training/optimization), the \gls{aerchemmip}, \gls{ramip}, \gls{rcp}, \gls{ssp}, and \gls{zecmip} scenarios are excluded from the optimization process. Exclusion does not by itself place a scenario outside the training distribution. For most, the individual forcing trajectories fall within the range spanned by the training pathways, and it is their combination that is out-of-distribution, while \gls{ssp}585, \gls{rcp}8.5, and \gls{zecmip} additionally exceed the optimization range in magnitude. Together, these scenarios test whether an emulator trained or optimized only on multi-forcing pathways reproduces the response to forcing structures absent from those optimization targets.

We generate optimized training emissions trajectories to maximize predictive skill on each set of scenarios individually, along with a configuration optimized over all scenarios simultaneously (Table \ref{tab:experiments}). Furthermore, because optimizing over all scenario sets at once introduces information leakage between the training and evaluation data, we perform an additional, independent evaluation. We use the optimized scenarios generated by our differentiable \gls{scm} as input to an \gls{emic}, training an emulator to reproduce the \gls{emic}'s zonal temperature response under an identical evaluation protocol. 

We first present the results of emulating \gls{gmst} from the \gls{scm}, followed by the results of emulating zonal temperatures from the \gls{emic}. Complete descriptions of emulator architecture, emissions scenarios, and evaluation protocol can be found in SI Appendices 2--5. Full numerical results for all experiments can be found in SI Appendix 7.

\textbf{SCM results: individual forcing agents.} We first focus on CO$_2$-only experiments, as the optimization results are qualitatively consistent across most agents (SI Appendix 7). Fig. \ref{fig:02_co2_example} provides an illustrative example of the optimization process when maximizing predictive skill for a high-warming emissions scenario (ScenarioMIP-CMIP7 Priority 1 \textit{H-ext}). While training an emulator on a naive, constant emissions time series (50 GtCO$_2$/yr) yields poor initial predictions (green dot-dash line, Fig. \ref{fig:02_co2_example}c), iterative updates to the training data drive the emulator's temperature predictions to near-perfect agreement with the \gls{scm}-projected targets. This convergence is robust across forcing agents, albeit at varying rates (Fig. \ref{fig:03_single_forcing}). The optimized emissions trajectory (Fig. \ref{fig:02_co2_example}b) differs structurally from the ground-truth emissions trajectory (Fig. \ref{fig:02_co2_example}a). While the optimization yields a pathway that shares some features with the ground truth, such as sign changes in the slope and concavity, it does not simply reconstruct \textit{H-ext}. This distinction suggests the optimization process (Fig. \ref{fig:01_optimization}) successfully isolates the physically salient features required for emulation, rather than memorizing a specific trajectory.

\begin{figure}[t]
    \centering
    \includegraphics[width=\linewidth]{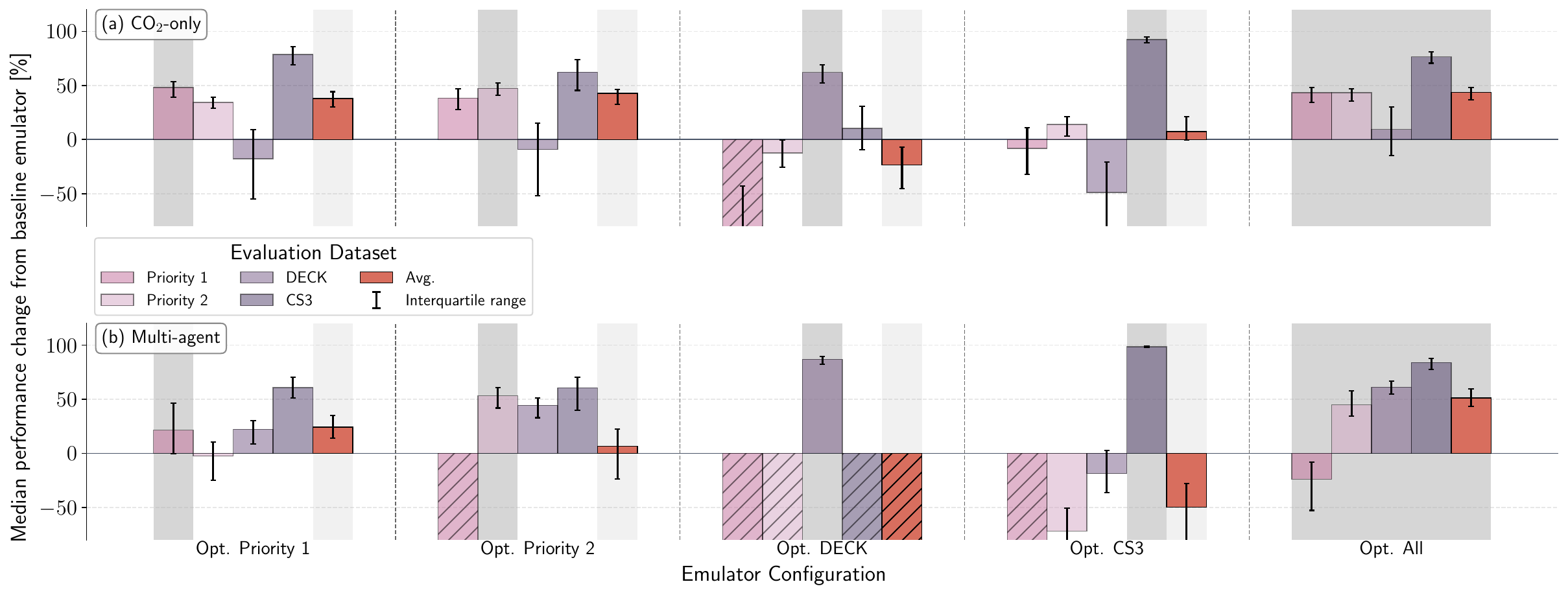}
    \caption{Performance of optimized emulators relative to the baseline emulator across several evaluation datasets. Change in predictive skill (\gls{nrmse}) from the baseline emulator for (a) CO$_2$-only and (b) multi-agent forcing experiments. Positive values indicate improved accuracy (reduced error). Emulators are evaluated across 50 independent stochastic runs. Bars represent the median performance across these seeds, where performance is calculated per seed as the mean \gls{nrmse} across all scenarios in the specified evaluation dataset. Vertical black lines indicate the interquartile range (25th to 75th percentile) of the seed distribution. Optimization targets include plausible policy pathways (ScenarioMIP-CMIP7 Priority 1 and 2, CS3), idealized forcing scenarios (DECK), and the full combined dataset (All). Gray-shaded bars indicate evaluations where data leakage is present; `Avg.' evaluations are shaded a lighter gray, since only part of `Avg.' overlaps with the optimization data. Hatched bars indicate a performance decrease exceeding y-axis limits. Full numerical results are provided in SI Appendix 7.}
    \label{fig:04_summary}
\end{figure}

Iterative optimization of training data yields higher predictive skill for individual forcing agents compared to the baseline emulator trained on standard socioeconomic scenarios (Fig. \ref{fig:03_single_forcing}). Because the baseline emulator is evaluated against its own training set (ScenarioMIP-CMIP7 Priority 1), this evaluation represents its theoretical error lower bound. Despite this, our optimized emulator achieves lower error (\gls{nrmse}) for all agents, crossing below the baseline emulator's error threshold (dark blue vs. dashed orange lines, Fig. \ref{fig:03_single_forcing}). This performance gap demonstrates that standard baseline scenarios are suboptimal for training, lacking the feature diversity necessary to capture all potential system behaviors. Transient increases or plateaus observed in the error trajectory (e.g., N$_2$O-, Sulfur-, and BC-only experiments) reflect the inherent trade-offs in multi-objective optimization, where aggregate skill gains across the full dataset may temporarily degrade performance on individual scenarios.

Fig. \ref{fig:04_summary}a summarizes the change in performance between the baseline and optimized emulator configurations for CO$_2$-only experiments, where positive values indicate improvement. Gray-shaded regions indicate evaluations where information leakage is present (i.e., the optimization target/baseline training data are contained within the evaluation set). Overall, optimizing for many of the plausible socioeconomic pathways (Opt. Priority 1 or 2), or for the combined dataset (Opt. All), consistently increases average emulator skill; the average performance improvement when optimizing for CS3 is driven by its in-objective evaluation and is not indicative of a robust emulator. Optimizing for all scenarios yields the largest average improvement (43--53\%), though this is expected as there is information leakage between the optimization targets and test scenarios. However, optimizing for the baseline Priority 1 scenarios yields similar average improvements (37--46\%) without information leakage to the other scenario sets. Notably, simultaneous optimization over all datasets yields performance gains across all evaluation datasets without overfitting to any specific scenario. While specialized optimization targets achieve the highest skill on their respective evaluation sets (e.g., optimizing for Priority 1 yields a 39--53\% increase when predicting Priority 1, compared to 34--48\% for the combined dataset), combined optimization ensures the emulator can generalize across scenario structures.

A clear trade-off emerges, however, regarding the idealized forcing scenarios (DECK). Optimizing for slowly varying socioeconomic pathways (Priority 1, Priority 2, or CS3) yields no improvement, or even degrades performance, on the idealized scenarios. Conversely, optimizing for the DECK reduces skill across all other datasets. This bifurcation stems from the idealized scenarios' unique forcing structure, specifically the abrupt quadrupling of CO$_2$ (\textit{abrupt-4xCO2}), which features a pulse-and-decline emissions trajectory driving rapid warming ($\mathcal{O}$(50 years) to reach 4$^\circ$C compared to $\mathcal{O}$(200 years) in other high-warming scenarios). While an idealized step forcing yields a skillful emulator for many data-driven approaches \cite{womack_theoretical_2026}, it acts as a statistical outlier during optimization. Minimizing emulator error on this shock without including the gentle gradients that characterize plausible socioeconomic emissions pathways, coupled with the idealized dataset's small sample size (two scenarios), promotes overfitting. In contrast, the comparably small CS3 dataset shares structural similarities with Priority 2, allowing for some extrapolation (3--21\%). Because the physical features required to emulate an emissions pulse conflict with those needed for more plausible emissions pathways, including the DECK in the combined optimization creates competing objective functions. However, it is necessary to include the idealized scenarios for generalization, as optimizing over all datasets results in the only majority-positive change in predictive skill on the abrupt scenario ($-15$ to $+30$\%), whereas optimizing only for plausible pathways yields no such improvement.

\begin{figure}[t]
    \centering
    \includegraphics[width=\linewidth]{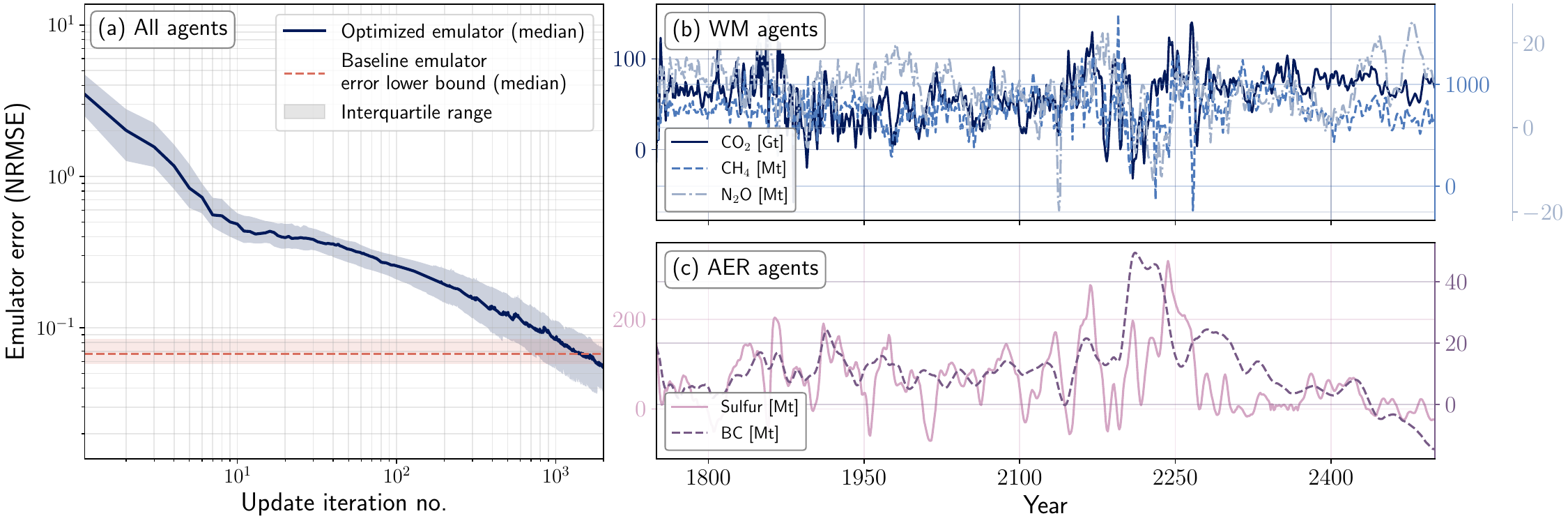}
    \caption{Emulator error and forcing trajectories for multi-forcing experiments. (a) Evolution of evaluation loss (\gls{nrmse}) when reproducing \gls{scm}-projected \gls{gmst} anomalies for plausible emissions pathways (ScenarioMIP Priority 1 with all forcing agents active). Emulators are evaluated across 50 independent stochastic runs. The solid dark blue line tracks the optimized emulator, while the dashed orange line indicates the baseline emulator's error lower bound (evaluated on its own training data). Shaded regions represent the interquartile range (25th to 75th percentiles) of the emulator errors. (b) Optimized emissions time series for well-mixed forcing agents (CO$_2$, CH$_4$, and N$_2$O). This panel displays one representative trajectory selected from the 50 stochastic runs. (c) Same as (b), but for aerosol agents (sulfur and black carbon). Full numerical results are provided in SI Appendix 7.}
    \label{fig:05_multi_forcing}
\end{figure}

\textbf{SCM results: multiple forcing agents.} Consistent with the single-agent results, the optimized emulator with all forcing agents active outperforms the baseline emulator when tested against standard socioeconomic projections (Priority 1), which represents the baseline emulator's theoretical error lower bound (Fig. \ref{fig:05_multi_forcing}a). The optimization process converges monotonically. Panels (b) and (c) of Fig. \ref{fig:05_multi_forcing} display the optimized time series for well-mixed and aerosol forcing agents. In the ground-truth plausible emissions pathways (Priority 1), all forcing agents follow highly correlated trajectories (e.g., CO$_2$ and CH$_4$ follow the same pattern of increase and decrease over time). This allows the baseline emulator to achieve high in-sample skill by learning aggregate forcing behavior rather than individual agent dynamics. While the optimized pathways we generate are structurally distinct from standard scenarios, they exhibit similar paired behavior, particularly CO$_2$ and CH$_4$, along with sulfur and BC. The well-mixed forcing agents additionally contain many high-frequency features, while the aerosol agents contain lower frequencies.

While optimizing for performance over individual datasets may lead to trade-offs in extrapolative skill (Fig. \ref{fig:04_summary}b), simultaneous optimization over the full scenario set yields performance gains across nearly every evaluation dataset. This result suggests that optimization isolates fundamental physical features independent of specific scenario structures; SI Appendix 6 demonstrates that a potentially infinite number of valid features exist, depending on the optimizer's initialization.  Incorporating a diverse set of scenarios during optimization can yield large improvements relative to the baseline for a specific target while retaining the ability to transfer across scenario sets. For example, when evaluated on the idealized DECK scenarios, the emulator optimized over the combined dataset (Opt. All) achieves a 61\% median improvement over the baseline while improving by a median of 53\% over all scenario sets. Opt. DECK exhibits a median decrease in skill of 780\% over all scenario sets, supporting the need for diverse optimization targets. Restricting the number of optimization targets degrades extrapolative performance further relative to the single-agent case. This is likely due to the increased complexity of emulating multiple agents and disaggregating their responses. Overfitting is most prevalent for the idealized DECK and plausible CS3 datasets, where small sample sizes (two scenarios each) and limited agent diversity (the DECK scenarios are CO$_2$-only) fail to adequately constrain the parameter space. 

\begin{figure}[p]
    \centering
    \includegraphics[width=\linewidth]{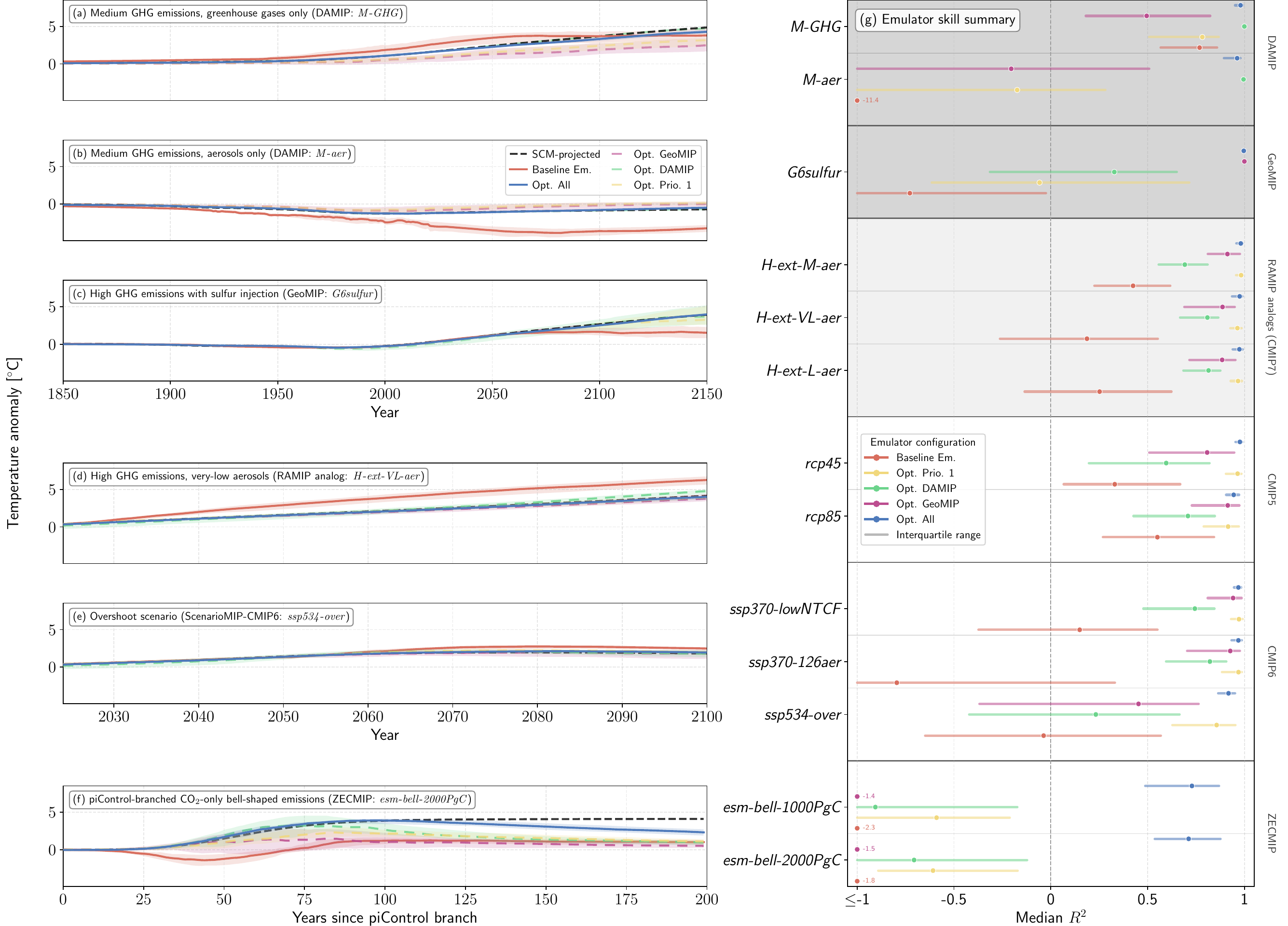}
    \caption{Emulator performance on structurally distinct forcing scenarios. (a)--(f) \gls{gmst} anomaly trajectories relative to pre-industrial conditions for (a, b) \gls{damip}, (c) \gls{geomip}, (d) \gls{ramip} analog, (e) ScenarioMIP-CMIP6, and (f) \gls{zecmip} scenarios. Lines compare \gls{scm} projections (dashed black), baseline predictions (solid orange), and the emulator optimized over all scenarios (solid blue) against emulators optimized for specific training subsets (dashed colored lines). Lines indicate median emulator performance over 50 stochastic seeds, and shaded regions indicate interquartile ranges. (g) Median $R^2$ across 50 training seeds for the baseline and optimized emulator configurations. Dark gray shading indicates scenarios are explicitly seen during optimization (e.g., the optimization set may include \textit{M-GHG}), while light gray shading indicates scenarios are implicitly seen during optimization (e.g., for \textit{H-ext-M-aer}, the optimization set may include \textit{H-ext} and \textit{M-aer} separately, but not the combined \textit{H-ext-M-aer}); no shading indicates scenarios that are fully out-of-distribution. The x-axis is clipped to $[-1,1]$; medians below $-1$ are pinned to the axis edge and annotated with their numerical value. Full numerical results are provided in SI Appendix 7.}
    \label{fig:06_individual_effects}
\end{figure}

Training an emulator on a scenario optimized for performance over all scenario types simultaneously (ScenarioMIP Priority 1 and 2, DECK, CS3, \gls{geomip}, and \gls{damip}) enables us to accurately reproduce both individual and aggregate forcing agent dynamics, correcting biases present in the baseline emulator (Fig. \ref{fig:06_individual_effects}); scenarios from CMIP5 (\glspl{rcp}), CMIP6 (\glspl{ssp} and \gls{aerchemmip}), and \gls{zecmip} are fully out-of-distribution and serve as an independent verification of this skill. As a first test, we evaluate the optimized emulators' ability to reproduce the isolated forcing and climate intervention subsets; emulating \gls{damip} and \gls{geomip} yields a median $R^2$ between 0.96 and 1 for Opt. All (Fig. \ref{fig:06_individual_effects}a--c, g). In contrast, emulators optimized for or trained on highly correlated aggregate emissions pathways (e.g., the Priority 1 scenarios) fail to generalize to these unseen datasets, exhibiting systematic errors. Neither the baseline emulator nor the Priority 1-optimized emulator accurately captures the distribution of warming and cooling effects between individual agents. For example, the baseline emulator systematically overestimates the cooling effect of sulfur. This failure is most evident when emulating \textit{G6sulfur}, a high-emissions climate intervention scenario that utilizes sulfur injection to limit warming. These emulators capture the aggregate trends prior to the geoengineering intervention but underestimate subsequent warming once sulfur injection begins.

The baseline emulator is also unable to reproduce the fully out-of-distribution scenarios, showing the lowest skill when emulating scenarios that require disaggregation of individual forcing agents (e.g., \textit{ssp370-lowNTCF}, \textit{ssp370-126aer}, and \textit{esm-bell-1000PgC}); the emulators optimized over only a subset of scenarios show a similar lack of skill. Only the emulator optimized on the full, diverse scenario set eliminates this bias, accurately predicting temperature anomalies across nearly the full range of individual and aggregate effects. One exception is the optimized emulator's performance on the CO$_2$-only \gls{zecmip} scenarios, where the emulator predicts a lower equilibrium temperature relative to the \gls{scm}, though this can likely be improved by modifying the selection of emulator features to include longer timescales or the history of emissions rates. Despite lower performance on the \textit{esm-bell} scenarios relative to the other out-of-distribution scenarios, the optimized emulator outperforms all other emulators by a wide margin (minimum median $R^2$ improvement from $-0.59$ to 0.73). This is attributable to the CO$_2$-only experiments in its optimization set, further underscoring the need for diversity in optimization targets.

\begin{figure}[t!]
    \centering
    \includegraphics[width=\linewidth]{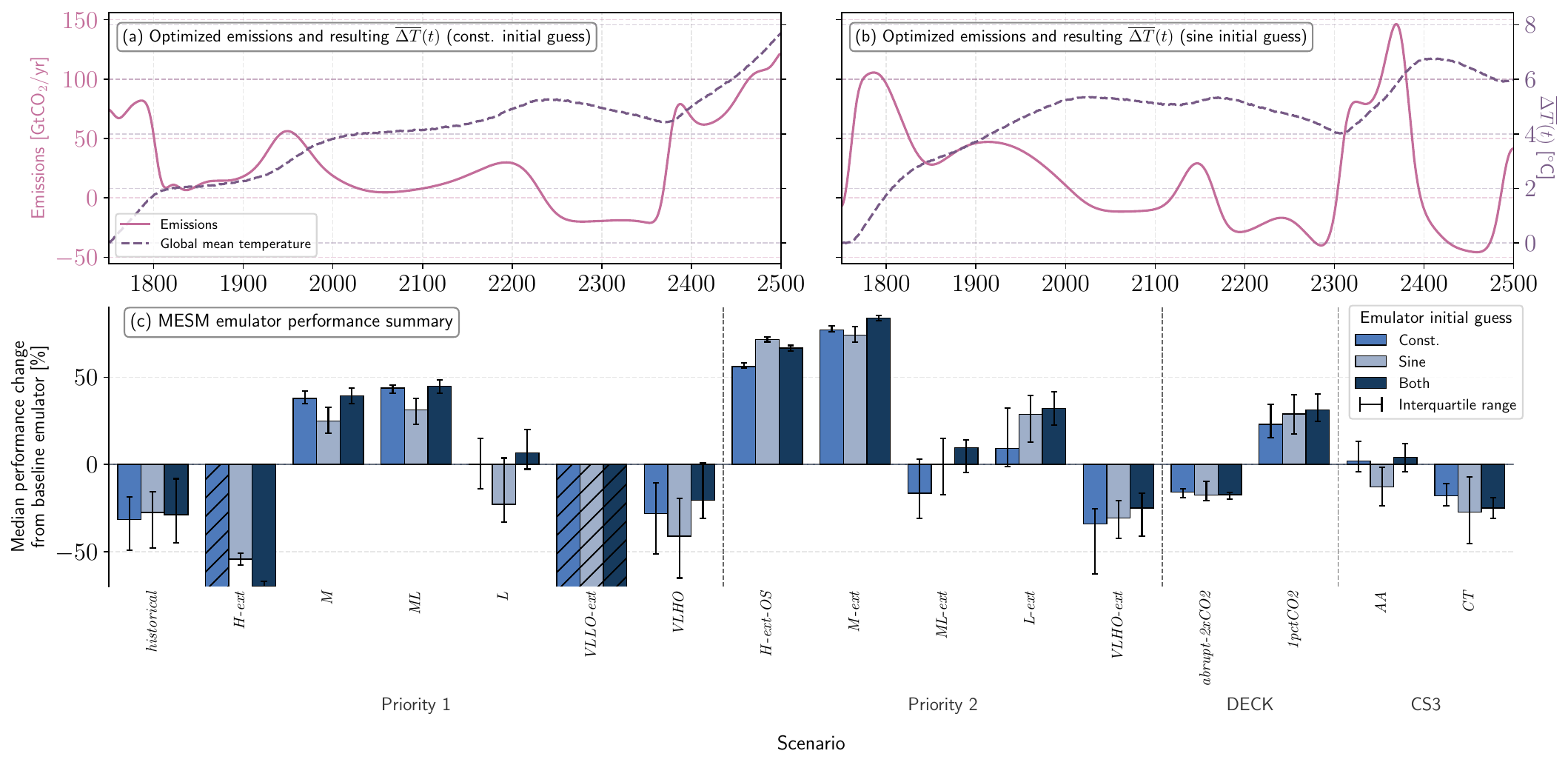}
    \caption{Training data and performance of optimized emulators relative to the baseline emulator across several evaluation datasets when emulating an intermediate-complexity climate model (\gls{mesm}). (a) Emissions and \gls{gmst} trajectories resulting from optimizing for predictive skill over all scenarios from a constant initial guess. This panel displays the highest-skill trajectory (tested on the \gls{scm}) selected from the 50 stochastic runs. (b) Same as (a), but from a sinusoidal initial guess. These emissions trajectories range from a minimum of $-21.10$ ($-45.74$) to a maximum of 121.08 (146.28) GtCO$_2$/yr for the constant (sinusoidal) initial guess. While \gls{gmst} is shown here for illustrative purposes, the baseline and optimized emulator configurations are trained to reproduce zonal temperature anomalies. (c) Change in predictive skill (\gls{nrmse}) from the baseline emulator for CO$_2$-only scenarios across plausible emissions pathways (ScenarioMIP Priority 1 and 2, CS3) and idealized scenarios (DECK). Positive values indicate improved accuracy (reduced error). Bars represent global average (latitude-weighted) performance. Const. and Sine refer to optimized scenarios starting from constant and sinusoidal initial guesses, respectively. Both indicates the emulator was trained on the scenarios from both sets of initial guesses. Hatched bars indicate a performance decrease exceeding y-axis limits. Full numerical results are provided in SI Appendix 7.}
    \label{fig:07_MESM_scens}
\end{figure}

\textbf{Intermediate-complexity model (MESM) results.} To validate our approach and demonstrate its scalability, we perform an independent evaluation using an intermediate-complexity climate model that outputs zonally averaged temperatures (\gls{mesm}). By utilizing the \gls{scm} from the previous section to generate optimized training scenarios that we then simulate with the intermediate-complexity model, we both verify that our optimized scenarios are useful for the more complex task of emulating zonal temperatures and conduct an experiment that prevents information leakage between the training and evaluation data. Due to operational constraints associated with running \gls{mesm} in emissions-driven mode, we limit our evaluation to CO$_2$-only scenarios. As before, we compare a baseline emulator trained on six plausible emissions scenarios (ScenarioMIP-CMIP7 Priority 1) against emulators trained on optimized scenarios, now using either one or two scenarios for training (derived from constant and sinusoidal initial guesses, Fig. \ref{fig:07_MESM_scens}a and b). Our results demonstrate that training on these optimized scenarios yields performance that matches or exceeds the six-scenario baseline emulator across both alternate policy projections (Priority 2) and idealized forcing scenarios (DECK). While the baseline emulator inherently retains the highest skill on its own training data (Priority 1), our optimized emulators demonstrate extrapolative improvements.

When emulating the intermediate-complexity model, the initial guess for the optimal emissions trajectory has a greater impact than in the case of the \gls{scm}; several initial guesses yield similar performance improvements over the baseline emulator for the \gls{scm} (SI Appendix 6). Optimizing from a sinusoidal initial guess generally yields higher predictive skill than optimizing from a constant initial guess, capturing a wider array of long-term physical dynamics; the choice of initial guess dictates which physical features the optimizer can isolate. Because the sinusoidal initial guess produces a trajectory with extended periods of decreasing and net-negative carbon emissions (Fig. \ref{fig:07_MESM_scens}b), we hypothesize it provides more informative features for extrapolating to new scenarios that exhibit these behaviors. Specifically, the centennial-scale oscillations present in the sinusoidal trajectory likely enable the optimization process to better constrain the characteristic timescales of the climate system; these temporal modes are required to accurately emulate delayed warming or cooling associated with physical processes like deep ocean heat uptake. This translates to increased skill, leading to a 26.8\% improvement over the baseline emulator on Priority 2 scenarios and a 5.6\% improvement on the idealized DECK. In contrast, optimizing from a constant initial guess produces a higher emissions trajectory (Fig. \ref{fig:07_MESM_scens}a) with a shorter period of net-negative emissions. As a result, though the constant-initialized emulator marginally outperforms the sinusoidal-initialized emulator on shorter, positive-emissions pathways (e.g., \textit{M}, \textit{ML}, \textit{L}, and \textit{VLHO}), it struggles to capture overshoot pathways like \textit{VLLO-ext} and \textit{H-ext-OS}, and exhibits mixed skill on the idealized DECK.

Including optimized scenarios from both initial guesses in the emulator training dataset (i.e., pooling the outputs of two separate optimization runs into a single expanded training dataset) yields, in almost every case, an improvement in skill over either individual configuration. Training on these two complementary scenarios drives a 32.6\% average improvement in extrapolative skill on Priority 2 scenarios, indicating that the combined dataset captures a broader spectrum of physical responses than either the six-scenario baseline or the individual optimized scenarios. However, combining multiple trajectories can occasionally lead to destructive interference that suppresses performance improvements. For example, performance on the Current Trends policy scenario (CS3 \textit{CT}) is lower than the average of the performance from training on the two scenarios separately. This likely occurs because the emulator attempts to average the distinct physical feedbacks triggered by the higher-warming constant initial guess and the more moderate sinusoidal initial guess, effectively interpolating to a non-physical intermediate state.  While this failure mode is restricted to this pathway and could potentially be resolved through scenario reweighting during optimization or joint optimization over multiple scenarios, our results indicate that optimizing across multiple initial guesses provides a pathway for training emulators that generalize across a wide range of future climates.

\section*{Discussion}

Generating maximally informative training data offers major utility for ML models of physical systems, particularly where data generation is computationally expensive. Our approach optimizes the training data for a climate emulator via a low-cost surrogate simple climate model, decoupling the computational cost of the optimization process from the run-time of full-scale Earth system models (\glspl{esm}). While training data for ML surrogate models are typically generated by costly numerical simulations \cite{ganti_datadriven_2020, zhang_machinelearningbased_2021}, our method produces optimal trajectories efficiently. This approach shares similarities with dataset distillation \cite{wang_dataset_2020, nguyen_dataset_2021, cazenavette_dataset_2022}, but differs fundamentally as we aim to generate a maximally informative dataset based on a specific target rather than identify salient features from an existing dataset. Our application to an intermediate-complexity climate model (\gls{mesm}) validates the scalability of this approach, showing increased predictive skill across structurally dissimilar scenarios.

As this study utilizes a simple multi-layer perceptron, our results provide a conservative estimate of the method's potential. While employing more complex sequence-based or attention-driven architectures would likely yield higher absolute predictive skill by more effectively capturing temporal dynamics, the simple neural network architecture highlights the relative benefit of the optimized training data themselves. As demonstrated by the sensitivity analysis to changes in the neural network architecture (SI Appendix 6, sinusoidal initial guess), the optimized time series generated by our approach are largely consistent across architectures. This suggests that the features extracted by our method are physically salient, rather than artifacts associated with a specific architecture. Further work across alternate ML architectures, physical domains, and systems with stronger nonlinearities is required to fully characterize our method's performance, but the method could apply more generally to any ML approach integrated with a differentiable synthetic data generation pipeline, highlighting its potential for the design of parsimonious training data.

Scenarios characterized by high structural diversity are better suited for both emulator training and understanding system behavior than baseline scenarios. The unconventional, rapidly varying emissions trajectories generated by our optimization process (Figs. \ref{fig:02_co2_example}, \ref{fig:05_multi_forcing}, \ref{fig:07_MESM_scens}, and S9) are highly informative for identifying a system's response, as suggested by system identification techniques \cite{kravitz_technical_2017}. To control for the effect of optimization separately from the structural diversity of a synthetic scenario, we include a comparison between the highest-performing emulator trained on a dynamically rich synthetic dataset and our optimized emulators in SI Appendix 7. The optimized emulator consistently outperforms the dynamically rich emulator across all configurations except the DECK, both with and without information leakage.

Our optimization results show that potentially many alternate choices for climate scenarios are more informative than the current choice of standard policy projections (ScenarioMIP), leading to higher predictive skill despite training on a smaller dataset. This is supported by Giani et al. \cite{giani_origin_2025} and Womack et al. \cite{womack_theoretical_2026}, who show that traditional high-warming emissions scenarios used for emulator training (e.g., \textit{SSP585}) can cause the Earth system's temperature response to reduce to a single timescale, rendering the emulator unable to learn the full system dynamics. Paired with the sensitivity analysis in SI Appendix 6, the generated trajectories illustrate that there is no single optimal scenario for training, but rather a family of optimal scenarios for a given application. This family is defined relative to an assumed budget for data generation, and the trajectories obtained in this work are optimal under the single-scenario constraint alone. A scenario that is optimal when only one simulation can be afforded need not belong to the optimal set when several can be run, since jointly optimized scenarios may each capture a subset of the system's dynamics rather than all of it. The same argument applies to scenario length, as the number of years available limits which timescales the optimizer can resolve. Determining the optimal allocation of a given budget across the number, length, and structure of scenarios is a question of experimental design under fixed cost, related to multi-fidelity modeling, and could be explored as future work. For example, training on multiple scenarios generated from distinct initial guesses (constant and sinusoidal) can yield an improvement in extrapolative skill over either individual configuration (Fig. \ref{fig:07_MESM_scens}).

Optimizing over all scenarios increases average emulator performance regardless of the number of forcing agents present (Fig. \ref{fig:04_summary}). This includes learning both individual and aggregate forcing behavior from a single scenario (Fig. \ref{fig:06_individual_effects}), with emulator performance validated on out-of-distribution scenarios. While previous emulation efforts have diagnosed forced responses from scenarios with a subset of forcing agents active \cite{schlesinger_geographical_2000, meehl_combinations_2004, wells_understanding_2023, dewey_aerogp_2025, sandstad_meteorv101_2025, bouabid_evidence_2026}, the ability to learn both individual and aggregate forcing behavior from a single scenario has not been explicitly demonstrated by previous work, illustrating the potential utility of our method.

We demonstrate the generalizability of our approach through the direct transfer of optimized scenarios between structurally distinct climate models (Fig. \ref{fig:07_MESM_scens}). Because transferability across model types is not guaranteed \textit{a priori}, the ability to use scenarios optimized exclusively on the simple climate model to train a skillful emulator for the intermediate-complexity model supports the practical utility of the method. One possible explanation for this cross-model skill is that an optimal scenario for learning the global response carries useful information for disaggregated regional responses, and future work could explore this in greater detail. Whereas the high computational cost of running thousands of simulations of a full-scale model prohibits us from directly identifying an optimal training scenario, this cross-model application indicates that a simple surrogate model may be sufficient for this purpose; future work can investigate simulating our optimized scenarios using a full-scale \gls{esm}. Although using the simple model as a surrogate requires optimizing over all scenarios simultaneously, which inherently introduces information leakage, our independent evaluation using the intermediate-complexity model confirms that we are able to successfully isolate salient physical features rather than merely overfitting to the evaluation metric.

Because our optimization procedure strictly requires a differentiable climate model, our work demonstrates the utility of differentiability across emulator training, calibration, and experimental design. First, we show that differentiability enables our approach to generate maximally informative training datasets. While backpropagation would be computationally intensive for a full-scale differentiable climate model \cite{moses_dj4earth_}, a modified version of this method could be used to inform online emulator training as a simulation is running, using the gradient to select the next data point that minimizes the emulator's loss. Second, we utilize the differentiability of our simple model to calibrate it to reproduce the median temperature response of the constrained, calibrated FaIR ensemble \cite{smith_faircalibrate_2024} without the expert intervention required by standard calibration techniques (e.g., minimizing the loss between observed and modeled climate statistics) \cite{kennedy_bayesian_2001, schneider_earth_2017, schneider_opinion_2024}; a differentiable model could be used to directly calibrate to observational data as well. Automatic differentiation accelerates this process and provides a systematic approach to calibration \cite{heimbach_efficient_2005, forget_ecco_2015, kochkov_neural_2024, davenport_jcm_2026a}. Finally, we use the model to generate the sulfur injection trajectory necessary to recreate the climate intervention scenario (\gls{geomip} \textit{G6sulfur}) via automatic differentiation. This allows us to compute the sensitivity of the output temperature to the sulfur trajectory, addressing the lack of consistent emissions protocols for such experiments \cite{kravitz_geoengineering_2015}.

Although the question of emulator interpretability is always present with nonlinear/black-box methods, our results highlight that the choice of training data plays a large role in an emulator's physical consistency. While our optimized emulator is not fully interpretable, its improved extrapolative capability may support the development of future emulators targeted toward interpretability. By successfully learning individual forcing effects and the full system response using only the scalar \gls{gmst} output from our optimized scenarios, we demonstrate a rigorous surrogate for \gls{esm} emulation where the availability of spatial information would likely simplify the separation of distinct forcing signatures.

However, there are trade-offs in this approach. Training on multiple structurally distinct scenarios can occasionally lead to destructive interference, as seen when the emulator attempts to average the physical behavior triggered by conflicting training regimes on the intermediate-complexity model's \textit{CT} scenario. Additionally, scenarios with extreme structural differences may have competing optimization goals, requiring more iterations to achieve high performance. Finally, our \gls{scm} gives an imperfect representation of \gls{mesm}'s temperature response (SI Appendix 4). It tracks \gls{mesm} to within $\sim$0.1$^\circ$C ($\sim$0.2$^\circ$C) along monotonically increasing (low-warming) pathways. However, it diverges in high-warming scenarios (peak \gls{gmst} anomaly $>$ 5$^\circ$C) that feature stabilizing or declining temperatures. There, \gls{mesm}'s warming persists while the \gls{scm} cools, with differences of up to 2.5$^\circ$C. This likely leaves room for improvement when designing an optimal scenario for a higher-complexity model. Future work can explore resolving these issues in two ways. Methodologically, ensemble learning concepts like boosting could sequentially generate optimal features missing from prior datasets \cite{friedman_stochastic_2002}, providing one route to extend our approach from a single scenario to a budget of several, where each new trajectory is optimized conditioned on what previous ones already capture. Physically, a two-step training procedure could separate system identification from optimization: (1) estimate intrinsic climate timescales through idealized experiments (e.g., \textit{abrupt-4xCO2}, as in Womack et al. \cite{womack_theoretical_2026}); (2) use sinusoids of those frequencies as initial guesses for our methodology, allowing the optimizer to find the remaining salient structures. 

As full-scale \glspl{esm} cannot keep pace with the ever-increasing demand for climate projections beyond \gls{cmip}, the popularity of climate emulators for scenario assessment continues to grow. While this study demonstrates the foundational theory and approach for generating optimal emulator training scenarios, fully realizing the utility of this method would require operational implementation to scale these results. This would involve applying it to a differentiable intermediate-complexity model (e.g., Davenport et al. \cite{davenport_jcm_2026a}) to evaluate how additional variables like precipitation alter the optimal emissions trajectory, and concurrently utilizing the trajectories derived in this work as forcing inputs for a full-scale \gls{esm}. Evaluating an emulator on these outputs would enable a direct comparison against a baseline emulator trained on standard policy projections.

Since previous work has shown that standard scenarios are suboptimal for emulator development \cite{womack_theoretical_2026}, modeling centers should consider dedicating resources to generate simulation data explicitly designed for machine learning. Optimizing a single scenario implicitly assumes the most restrictive computational budget a modeling center is likely to face. Under this constraint, a single trajectory must fully characterize the system's response. A larger budget would relax this constraint, allowing several trajectories to be optimized simultaneously such that each captures dynamics the others leave out; we expect this to yield greater performance gains and extrapolative improvements than we report here. The gains from pooling two independently optimized scenarios support this expectation (Fig. \ref{fig:07_MESM_scens}), while the destructive interference observed on the \textit{CT} scenario indicates that joint optimization, rather than scenario pooling, is the appropriate path forward. Regardless of budget, the inherent uncertainty of future socioeconomic pathways (e.g., CMIP8 and beyond) requires training emulators to capture the broadest possible range of climate dynamics; the optimization presented in this work provides one structured approach to achieve this scenario diversity. Because these trajectories diverge significantly from standard model intercomparison protocols, they effectively force models into regimes outside their typical tuning. These stress tests can offer utility beyond emulator training by quantifying model uncertainties under out-of-distribution forcings, potentially providing information on where model structural assumptions break down. Establishing a formal intercomparison project for emulator development beyond the FASTMIP protocol would benefit both the climate modeling and impacts communities \cite{windisch_advancing_2026}. Such an initiative would produce robust emulators capable of generating large, impact-relevant ensembles in a fraction of the time, ultimately freeing computational resources to focus full-scale models on frontier Earth system science.

\section*{Materials and Methods}

\textbf{1. Training data optimization.} We frame the generation of training data as a bi-level optimization problem (Fig. \ref{fig:01_optimization}); we outline our procedure here and include more detail in SI Appendix 1. Our objective is to find a specific set of training emissions ($\mathbf{U}_\text{train}$) that minimizes the error of an emulator trained on those data when tested against a target set. This problem consists of an implicit inner level (training the emulator parameters $\theta$) and an explicit outer level (updating the training emissions). The optimization objective is given mathematically as 
\begin{equation*}
    \operatorname*{argmin}_{\mathbf{U}_\text{train}} \mathcal{L}_{\text{test}}(\mathbf{U}_\text{train}, \, \theta_\text{train}, \, D_{\text{test}}),
\end{equation*}
where $\theta_\text{train}$ represents the parameters of the emulator after training on the data generated by $\mathbf{U}_\text{train}$, and $D_{\text{test}}$ is a test dataset held constant during optimization. The dimensions of $\mathbf{U}_\text{train}$, namely the number of scenarios and the number of years in each, are fixed prior to optimization and are not themselves optimized. These dimensions assume a budget for data generation, so any solution we obtain is conditional on that budget. Except where noted for the intermediate-complexity model, we optimize a single trajectory of fixed length, corresponding to a budget of one simulation.

\textbf{1.1 Inner level (emulator training)}: The inner level consists of training an emulator to map from emissions to temperature anomalies. We construct training features ($\mathbf{X}_\text{train}$) from our emissions time series ($\mathbf{U}_\text{train}$). We then force the \gls{scm} with $\mathbf{U}_\text{train}$ to generate the corresponding \gls{gmst} anomalies ($\mathbf{y}_\text{train}$), which serve as ground-truth targets. The emulator is trained via stochastic gradient descent (SGD) to minimize the mean squared error between its predictions and $\mathbf{y}_\text{train}$, resulting in optimized network weights ($\theta$).

\textbf{1.2 Outer level (emissions update)}: The outer level tests the performance of the trained emulator on $D_{\text{test}}$. We quantify test performance using scenario-length-weighted \gls{nrmse} ($\mathcal{L}_\text{test}$); length-weighting prevents short scenarios from being overrepresented during optimization. To update $\mathbf{U}_\text{train}$ to minimize $\mathcal{L}_\text{test}$, we utilize automatic differentiation to efficiently calculate the gradient $\nabla_{\mathbf{U}_\text{train}} \mathcal{L}_\text{test}$ by backpropagating through the testing, training, and data generation processes. We then apply these updates via an SGD optimizer with momentum \cite{liu_improved_2020}. A complete breakdown of the chain-rule expansion of our procedure and the corresponding pseudocode is provided in SI Appendix 1.2.

\textbf{2. Simple and intermediate-complexity climate models.} To enable our optimization procedure, we present a differentiable implementation of an \gls{scm} based on the \gls{fair} \gls{scm} \cite{leach_fairv200_2021}. Implemented in JAX, this model leverages automatic differentiation for efficient gradient-based calibration while retaining the core structural components of \gls{fair}. We use a three-box impulse response model to calculate \gls{gmst} anomaly time series based on total effective radiative forcing from five forcing agents: CO$_2$, CH$_4$, N$_2$O, sulfur, and black carbon; a full description of the model and its calibration can be found in SI Appendix 2.

To more rigorously test the suitability of our optimized scenarios for training emulators of more sophisticated models than our \gls{scm}, we utilize \gls{mesm}, an \gls{emic} that includes a two-dimensional, zonally averaged atmospheric model with interactive chemistry coupled to a zonally averaged land model and an anomaly-diffusing ocean model; see Sokolov et al. \cite{sokolov_description_2018} for a full description. As \gls{mesm} is not differentiable, we utilize outputs from the optimization procedure using our differentiable \gls{scm} as inputs to \gls{mesm}, simulating the zonal temperature response to these emissions. We additionally simulate the scenarios outlined in SI Appendix 5; all \gls{mesm} simulations are run as a thirty-member initial-condition ensemble.

\textbf{3. Neural network emulator.} We implement a neural network emulator to predict temperature from emissions for both climate models considered in this work. We emulate \gls{gmst} from our \gls{scm} and ensemble-average zonal temperatures from \gls{mesm}. As this work focuses on the impact of training data, rather than emulator architecture (i.e., emulator structure and feature design), on predictive skill, we use a multi-layer perceptron, the simplest possible neural network; improvements in predictive skill are likely possible with more advanced architectures. We train several emulator configurations for each climate model: a baseline emulator trained on the Priority 1 scenarios from ScenarioMIP-CMIP7, along with one emulator for each set of optimized training data as described in SI Appendix 5. While both baseline and optimized emulators utilize a fixed number of gradient steps during training, hyperparameters for each emulator configuration are tuned independently to maximize individual emulator skill. A full description of the hyperparameters and emulator architectures used for each climate model can be found in SI Appendix 3.

\textbf{4. AI use disclosure.} The authors declare the use of generative AI in the research and writing process. According to the GAIDeT taxonomy (2025), the following tasks were delegated to GAI tools under full human supervision:

\begin{itemize}
    \item Code generation
    \item Code optimization
    \item Proofreading and editing
\end{itemize}

The GAI tools used were: Gemini 3.1 Pro along with Claude Sonnet 5 and Opus 5.
Responsibility for the final manuscript lies entirely with the authors.
GAI tools are not listed as authors and do not bear responsibility for the final outcomes.

\section*{Data, Materials, and Software Availability}
The codebase accompanying this work, along with all data required to reproduce this work, is made publicly available on \href{https://doi.org/10.5281/zenodo.22875971}{Zenodo}\footnote{\url{https://doi.org/10.5281/zenodo.22875971}}.

\section*{Acknowledgements}
This research was part of the Bringing Computation to the Climate Challenge (BC3) project and supported by Schmidt Sciences through the MIT Grand Challenges. Development of the MESM model used in the analysis is supported by an international consortium of government, industry, and foundation sponsors of the MIT Center for Sustainability Science and Strategy. \href{https://cs3.mit.edu/sponsors/current}{See here for a complete list}\footnote{\url{https://cs3.mit.edu/sponsors/current}}. We acknowledge the MIT \textit{Svante} cluster supported by the Center for Sustainability Science and Strategy for computing resources. We also acknowledge the MIT Office of Research Computing and Data for providing high-performance computing resources that have contributed to the research results reported within this paper. We are grateful to the entire BC3 team, who provided insightful feedback and discussions about this work. We would also like to thank Chris Smith for his advice on running \gls{fair}, along with Claudia Tebaldi, Raffaele Ferrari, Will Chapman, David Darmofal, and Darya Guettler for their feedback throughout the writing of this manuscript. Finally, we would like to thank the two anonymous reviewers for their thorough comments and time spent reviewing this manuscript.

\bibliographystyle{unsrtnat}
\bibliography{references}

\end{document}